\documentclass[12pt,oneside]{amsart}
\usepackage[margin=1in]{geometry}
\usepackage{amsmath,amsthm,amssymb,amsfonts}
\usepackage{graphicx}
\usepackage{latexsym}
\usepackage[T2A]{fontenc}
\usepackage[utf8]{inputenc}
\usepackage[russian,english]{babel}
\usepackage{natbib}
\usepackage{ragged2e}
\usepackage{setspace}
\usepackage{xcolor}
\usepackage{csquotes}

\theoremstyle{definition}
 \newtheorem{theorem}{Theorem}
 \newtheorem{example}{Example}

\begin{document}

\thispagestyle{empty}

\title{Measuring Majority Power and Veto Power of Voting Rules}

\author{Aleksei Y. Kondratev}

\author{Alexander S. Nesterov}

\thanks{\phantom{ } Aleksei Y. Kondratev --- National Research University Higher School of Economics, 16, Soyuza Pechatnikov st., St.~Petersburg, 190121, Russia; Institute for Regional Economic Studies RAS, 38, Serpuhovskaya st., St.~Petersburg, 190013, Russia --- akondratev@hse.ru --- http://orcid.org/0000-0002-8424-8198\\
\phantom{ } Alexander S. Nesterov --- National Research University Higher School of Economics, 16, Soyuza Pechatnikov st., St.~Petersburg, 190121, Russia --- asnesterov@hse.ru --- http://orcid.org/0000-0002-9143-2938\\
\phantom{ } Support from the Basic Research Program of the National Research University Higher School of Economics is gratefully acknowledged. Kondratev is partially supported by the Russian Foundation for Basic Research via the project no.~18-31-00055. Nesterov is partially supported by the Grant 19-01-00762 of the Russian Foundation for Basic Research.\\
\phantom{ } We are grateful to Richard Ericson, Shmuel Nitzan and Dan Felsenthal for their encouraging and valuable feedback. We thank participants of the 7th International Workshop on Computational Social Choice, the 14th Meeting of the Society for Social Choice and Welfare, the 10th Lisbon Meetings in Game Theory and Applications for their helpful comments; we also thank Elena Yanovskaya, and other our colleagues from the HSE St.~Petersburg Game theory lab for their suggestions and support.\\
\phantom{ } Preliminary versions of this paper have circulated under the titles ``Weak Mutual Majority Criterion for Voting Rules,'' ``Measuring Majority Tyranny: Axiomatic Approach.''}

\date{03 June 2019}                     

\begin{abstract}
We study voting rules with respect to how they allow or limit a majority from dominating minorities: whether a voting rule makes a majority powerful, and whether minorities can veto the candidates they do not prefer. For a given voting rule, the minimal share of voters that guarantees a victory to one of their most preferred candidates is the measure of majority power, and the minimal share of voters that allows them to veto each of their least preferred candidates is the measure of veto power. We find tight bounds on these minimal shares for voting rules that are popular in the literature and in real elections. We order these rules according to majority power and veto power. The instant-runoff voting has both the highest majority power and the highest veto power and the plurality rule has the lowest. In general, the higher the majority power of a voting rule is, the higher its veto power. The three exceptions are: voting with proportional veto power, Black's rule, and Borda rule, which have a relatively low level of majority power and a high level of veto power and thus provide minority protection. Our results can shed light on how voting rules provide different incentives for voter participation and candidate nomination.

\medskip
\medskip


\medskip

\end{abstract}

\maketitle

\newpage

\thispagestyle{empty}

\section{Introduction}\nopagebreak
\label{Introduction}

\onehalfspacing

Majority tyranny has been a buzzword for centuries and can be traced back to the ancient Greek ochlocracy. A more modern yet classical reference is the work of James Madison:

\begin{displayquote}
If a majority be united by a common interest, the rights of the minority will be insecure. (Federalist 51.)
\end{displayquote}

In this paper we propose a simple way to quantitatively measure the robustness of a voting rule to majority tyranny or, more mildly, majority power, that is, the extent to which this rule allows a majority to dictate the outcome of an election regardless of the minorities' opinion and voting strategy.

Consider the following illustrative example presented in Table~\ref{profile_Donald}. Let there be five candidates: Bernie, Donald, Hillary, John and Ted, and let the voters have one of the five rankings of the candidates, with the top row giving the share of these voters in the population. Here, as also throughout the paper, the total number of voters is arbitrary.

\begin{table}[ht!]
\caption{Preference profile}
\label{profile_Donald}
\begin{tabular}{|c|c|c|c|c|c|}
\hline
share of voters & 22\% & 21\% & 18\% & 19\% & 20\% \\
\hline
1st candidate & Hillary & Donald & \textcolor{blue}{\underline{John}} & \textcolor{blue}{\underline{Ted}} & \textcolor{blue}{\underline{Bernie}} \\
2nd candidate & John & John & \textcolor{blue}{\underline{Ted}} & \textcolor{blue}{\underline{Bernie}} & \textcolor{blue}{\underline{John}} \\
3rd candidate & Bernie & Ted & \textcolor{blue}{\underline{Bernie}} & \textcolor{blue}{\underline{John}} & \textcolor{blue}{\underline{Ted}} \\
4th candidate & Ted & Bernie & Donald & Donald & Hillary \\
5th candidate & Donald & \textit{\textcolor{red}{Hillary}} & \textit{\textcolor{red}{Hillary}} & \textit{\textcolor{red}{Hillary}} & Donald \\
\hline
\end{tabular}
\vspace{0.2cm}
\justify
\footnotesize{\emph{Notes}: All characters and numbers in this example are fictitious. }
\end{table}

Let us look at the voters in the last three columns in Table~\ref{profile_Donald}. These voters constitute a \textbf{mutual majority} of 57\% as they prefer the same subset of candidates (Bernie, John, and Ted) over all other candidates. Depending on the voting rule, this 57\% may be enough to guarantee that one of the three candidates will win. For example, it is enough for the \textbf{instant-runoff voting}. According to this rule, from each ballot we iteratively delete the candidate with the fewest top positions. John is deleted first, followed by Bernie, then Donald, then Hillary, and the winner is Ted.\footnote{The instant-runoff is a special case of the single transferable vote (STV) when we select a single winner.} In contrast, the plurality rule makes Hillary the winner, and the plurality with runoff deletes each candidate except for the two with the most top positions (Donald and Hillary) and thus makes Donald the winner.

Formally, a voting rule satisfies the \textbf{$(q,k)$-majority criterion} if whenever a group of $k$ candidates get top $k$ positions among a qualified mutual majority of more than $q$ voters then the rule must select one of these $k$ candidates. For a given voting rule, the $(q,k)$-majority criterion measures its majority power: the lower the quota $q$, the more the rule empowers the majority.

The $(q,k)$-majority criterion subsumes few criteria known in the literature. The \textbf{majority criterion} requires that a candidate top-ranked by more than half of the voters is declared the winner. This criterion is equivalent to $(q,k)$-majority criterion with $k=1$ and a fixed quota $q=1/2$. When we consider a mutual majority that top-ranks some $k$ candidates then the \textbf{mutual majority criterion} requires that one of these $k$ candidates wins. This criterion is equivalent to $(q,k)$-majority criterion with an arbitrary $k$ and a fixed quota $q=1/2$.

\begin{figure}
\begin{center}
\includegraphics[width=14cm]{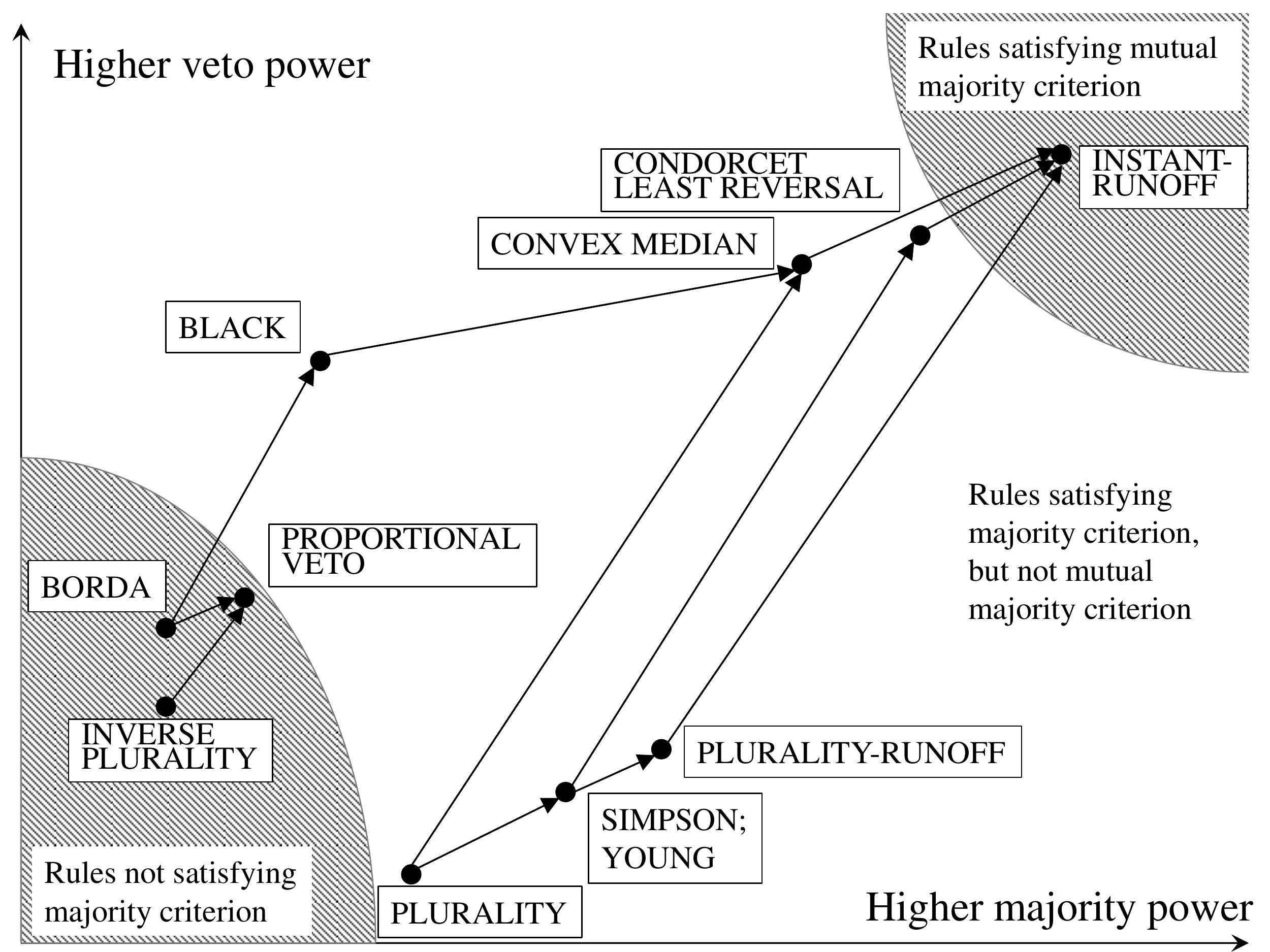}
\end{center}
\caption{Majority power and veto power: partial comparison of voting rules}
\label{partial-order}
\vspace{0.2cm}
\justify
\footnotesize{\emph{Notes}: Black dots represent voting rules. Arrows go from rules that have lower majority power and lower veto power to the rules that have higher power in an arbitrary setting. The exact positioning of the rules is based on authors' interpretation of tables \ref{table-k-1-2-3-4-5}, \ref{table-m-3-4}, \ref{table-veto}, \ref{table-veto-2} for different settings. Simpson's rule and Young's rule always have the same majority power and veto power.}
\end{figure}

The previous literature studied majority power by partitioning the voting rules into three categories: (1) rules that do not satisfy the majority criterion, (2) rules that satisfy the majority criterion but not the mutual majority criterion, and (3) rules that satisfy the mutual majority criterion (see Fig.~\ref{partial-order}).
	
Category (1) has the smallest majority power as these rules do not guarantee a majority that their $k = 1$ top candidate wins. This category includes the proportional veto core \citep{Moulin81,Moulin82,Moulin83book} and positional scoring rules like the Borda rule \citep{BaharadNitzan02,Nitzan09}. In contrast, category (3) has the largest majority power, as even a simple majority ($q\geq 1/2$) is enough to vote through one of its top-ranked candidates; a review of these results can be found, for example, in \citet{Tideman06}. However, these criteria are black-or-white and do not allow a finer analysis of majority power. Our quantitative criteria fill this gap.

Among all voting rules of interest, the most important are perhaps the plurality rule and the plurality with runoff rule. Together with the instant-runoff these voting rules are most widespread in political elections around the world.\footnote{A version of plurality with runoff -- \emph{two-round system} -- is used for presidential elections in France and Russia. The US presidential election system with primaries also resembles the plurality with runoff rule given the dominant positions of the two political parties. The instant-runoff voting is currently used in parliamentary elections in Australia and presidential elections in India and Ireland. According to the Center of Voting and Democracy (fairvote.org, \citeyear{Fairvote}) the instant-runoff and plurality with runoff rules have the highest prospects for adoption in the US. In the UK, a 2011 referendum proposing a switch from the plurality rule to the instant-runoff voting lost when almost 68\% voted \emph{No}.} Interestingly, these three voting rules are comparable in terms of majority power in an arbitrary setting. The instant-runoff voting makes the majority extremely powerful and has a constant quota $q=1/2$ as it satisfies the mutual majority criterion. We show that the plurality with runoff makes the majority less powerful, $q=\max\{k/(k+2),1/2\}$, while the plurality rule empowers the majority the least among these three rules, $q=k/(k+1)$. 

In our example presented in Table~\ref{profile_Donald}, $k=3$ and thus in order to guarantee a victory to one of the preferred candidates the share of voters must be more than 60\% under the plurality with runoff and more than 75\% under the plurality rule. In the example, however, the share is only 57\%, which is not enough to guarantee a victory for either John, Bernie or Ted under these two voting rules.

As an additional simple illustration of the quotas, consider how these three voting rules give different incentives for \textbf{candidate nomination}. A leading party (or a coalition) that has the support of at least half the voters decides whether to nominate two candidates in a general election or run primaries and nominate a single candidate. Under the plurality with runoff, the party is safe to forgo primaries and nominate both candidates.\footnote{For example, this is done in non-partisan blanket primaries in the US, which is a version of plurality with runoff.} Under the instant-runoff voting this is also the case; moreover, this party can safely nominate more than two candidates. But under the plurality rule, unless the party has the support of at least 2/3 of voters, it has to run primaries and can only nominate a single candidate.

The summary of our main results (see Theorem \ref{theorem_plurality} to \ref{theorem_inverse_plurality}) is presented in Fig.~\ref{partial-order}. We focus on the voting rules that are popular in the literature and do not satisfy the mutual majority criterion, finding the minimal size of the qualified mutual majority $q$ for each of these rules. In Fig.~\ref{partial-order} the voting rules are ranked in increasing order of majority power: each arrow goes from a rule with a larger minimal quota to a rule with a smaller minimal quota.\footnote{More specifically, the comparison is made for a given setting: for each given total number of candidates~$m$ and each given number of preferred candidates~$k$ we find the minimal size of qualified mutual majority $q(k,m)$.} Interestingly, while for a small number of preferred candidates~$k$ we get only \emph{partial} order, as shown in Fig~\ref{partial-order}, whenever $k>4$ and the total number of candidates~$m$ is arbitrary we get a \emph{complete} majority power order over the rules (see the x-axis in Fig.~\ref{partial-order} and Table~\ref{table-k-1-2-3-4-5} for details).

Our analysis allows us to ask a question that is dual to majority power -- the question of \textbf{veto power}. Imagine there is a group of voters that dislike a group of $l$ candidates and give them the lowest ranks in an arbitrary order. In our example presented in Table~\ref{profile_Donald} we have two such groups: 58\% of voters dislike Hillary ($l=1$), while 57\% of voters dislike Hillary and Donald ($l=2$). The question is: are these groups large enough to prevent each of these $l$ candidates from winning? 

Formally, a voting rule satisfies the \textbf{($q,l$)-veto criterion} if any group of a size larger than $q$ can always veto each of these $l$ candidates. For a given voting rule, the $(q,l)$-veto criterion measures its veto power: the lower the quota $q$ is, the higher its veto power.

The veto criterion generalizes the majority loser criterion known in the literature. The \textbf{majority loser criterion} requires that when more than half the voters give the same candidate the lowest rank then this candidate does not win. This criterion is equivalent to the $(q,l)$-veto criterion with $l=1$ and $q=1/2$.

Based on the $(q,l)$-veto criterion, we get \emph{the same} partial order as in the case of the $(q,k)$-majority criterion (Fig.~\ref{partial-order}). For example, the instant-runoff voting gives the highest veto power, the plurality rule gives the lowest veto power, and the plurality with runoff rule is in-between. This is not surprising due to duality: for a given number of candidates $m$, a group that vetoes its $l$ least preferred candidates at the same time guarantees that one of their $k=m-l$ most preferred candidates wins.\footnote{Due to this duality, the $(q,k)$-majority criterion and the $(q,l)$-veto criterion can together be referred to as the \emph{qualified mutual majority criterion}.}

However, when we consider the $(q,l)$-veto criterion for an arbitrary total number of candidates~$m$ then the order of the voting rules is \emph{different} from the order for the $(q,k)$-majority criterion. Whenever $l>2$ the \emph{complete} order based on veto power is presented in Table~\ref{table-veto} and shown on the y-axis in Fig.~\ref{partial-order}.

Based on our two criteria, the instant-runoff voting is the most powerful rule both from the majority power viewpoint and the veto power viewpoint. Since the votes are transferable, a simple majority of voters that like the same candidates can ensure the win of one of these candidates. Similarly, if a simple majority dislikes the same candidates then none of these candidates wins. Thus, the instant-runoff voting \emph{protects simple majorities}.
At the other extreme, the plurality rule appears to have a relatively low level of majority power and veto power. 

Perhaps surprisingly, the tradeoff between majority power and veto power has three exceptions. Among all voting rules, the \textbf{proportional veto core},\footnote{In this rule each group of voters can veto the share of candidates that is approximately the same as the share of this group. The rule selects the candidates that have not been vetoed by any group.} \textbf{Black's rule},\footnote{Black's rule selects a Condorcet winner (aka pairwise majority winner) if it exists and a Borda winner otherwise.} and the \textbf{Borda rule} provide a balanced combination of properties: relatively low majority power and high veto power. Thus, these three rules are the best to \emph{protect minorities}. In section~\ref{Conclusions} we discuss why these three rules are exceptional in our framework.

Throughout the paper we ignore strategic issues and treat the set of candidates, the set of voters, and their preferences as fixed.\footnote{The literature on strategic voting is prolific; see e.g., \citet{KondratevMazalov19} and references therein.} Nevertheless, our two criteria and the results have another interpretation: they can shed light on how voting rules differ in their incentives for \textbf{voter participation}. Consider an election where a mutual majority of size $q$ top-ranks some $k$ candidates. If the minimal quota of a given rule is lower than $q$ then minorities cannot do much. At most, minorities can influence which of the top-ranked $k$ candidates is selected,\footnote{In fact, they might be unable to even do this. In our example presented in Table~\ref{profile_Donald}, under the instant-runoff, the 43\% minority prefers John but John is deleted first.} but any other candidate has zero chance of winning regardless of the minority vote. This can discourage minorities from participating, thereby making the relative size of the mutual majority larger and minorities smaller and weaker, causing positive feedback.

In contrast, when the minimal quota is larger than $q$ then minorities have a larger influence on the election and hence have stronger incentives to participate. If it causes minorities to show up more then their relative size becomes larger again, causing positive feedback.

A similar participation argument can be  deduced from the veto criterion. For instance, in a given election let the ruling party nominate $l$ candidates (when $l=1$ this candidate can be the incumbent), and let the opposition nominate their own candidates. The opposition supporters do not necessarily agree on their most preferred candidate, but they agree that the nominees of the ruling party are the worst. Then, a voting rule with low veto power (i.e., high minimal quota) discourages opposition supporters from participating, while a rule with high veto power (i.e., low minimal quota) provides stronger incentives to participate.

The two participation arguments have opposite implications. The veto power argument predicts that the voting rules with low quotas encourage the participation of opposition supporters who dislike the same candidates, while the majority power argument suggests that such rules discourage the minorities from struggling with a mutual majority. 

As an example, consider the case of the instant-runoff voting. The common opinion of the social choice literature is that the instant-runoff voting promotes participation. Indeed, when minorities face a larger group of voters with a single preferred candidate then the instant-runoff allows the minorities to transfer their votes instead of wasting them \citep{Tideman95,Zwicker16}. However, if minorities face a mutual majority with more than one preferred candidate then the instant-runoff works in favor of this majority.

Overall, we abstain from normative judgment and cannot say which voting rule is the best based on the two criteria that we propose. These two criteria are instruments that should be used at the discretion of a mechanism designer, as such decisions always involve tradeoffs. We arrive at some of these tradeoffs when we study the compatibility of the $(q,k)$-majority criterion (and the $(q,l)$-veto criterion) with other axioms; these results are presented in subsection~\ref{ssec_tradeoffs}.  

The paper proceeds as follows. Section~\ref{Model} presents the model and the definitions of the voting rules. In section~\ref{Main} we analyze our two criteria for the voting rules and the relevant tradeoffs. A non-technical reader who understands the two criteria may safely skip these two sections, only taking a look at tables \ref{table-k-1-2-3-4-5}, \ref{table-m-3-4}, \ref{table-veto}, \ref{table-veto-2}, and proceed to section~\ref{Conclusions} which concludes with a discussion of the results and the open questions. The proofs are presented in the Appendix.


\section{Model}\nopagebreak
\label{Model}

\subsection{Voting Problem}

This subsection introduces the standard voting problem and the main criteria for voting rules.

Consider a \textbf{voting problem} where $n \geq 1$ voters $I=\{1,\ldots, n\}$ select one winner among $m\geq 1$ candidates (alternatives) $A=\{ a_1, \ldots, a_m\}$. Let $L(A)$ be the set of linear orders (complete, transitive, and antisymmetric binary relations) on the set of candidates~$A$.

Each voter $i \in I$ is endowed with a \textbf{preference relation} $\succ_i \in L(A)$. (Voter~$i$ prefers $a$ to~$b$ when $a\succ_i b$.)

Preference relation $\succ_i$ corresponds to a unique ranking bijection $R_i : A \rightarrow \{1,\ldots, m\}$, where $R_i^a$ is the relative rank that voter $i$ gives to candidate~$a$,
\begin{equation*}
R_i^a = |\{ b\in A : b\succ_i a \}| + 1, \quad a\in A, \quad i\in
\{1,\ldots, n\}.
\end{equation*}

The collection of the individual preferences $\succ$ = $(\succ_1, \ldots,\succ_n)$ $\in {L(A)}^n$ as well as the corresponding ranks $(R_1, \ldots, R_n)$ are referred to as the \textbf{preference profile}. (There exist $m!$ different linear orders and $(m!)^{n}$ different profiles.)

\begin{example}\label{example_main}
Table~\ref{profile_main} provides an example of a preference profile for $n=100$ voters over $m=4$ candidates. Here, voters are assumed to be anonymous, which allows us to group voters with the same individual preferences. Each column represents some group of voters, with the number of voters in the group in the top row; the candidates are listed below (starting from the most preferred candidate) according to the preference of the group.
\end{example}

\begin{table}[ht!]
\caption{Preference profile, tournament matrix, and positional matrix}
\label{profile_main}
\begin{tabular}{|c|c|c|c|}
\hline
29 & 28 & 22 & 21  \\
\hline
a & b & c & c  \\
b & a & d & d  \\
c & c & a & b  \\
d & d & b & a  \\
\hline
\end{tabular}
\quad
\begin{tabular}{|c|c|c|c|c|c|c|}
\hline
  & a & b & c & d   \\
\hline
a &   &  {\bf 51} & {\bf 57} & {\bf 57}  \\
\hline
b &  49 &   & {\bf 57} & {\bf 57}   \\
\hline
c &  43 & 43 &  & {\bf 100}  \\
\hline
d & 43 & 43 &  0 &    \\
\hline
\end{tabular}
\quad
\begin{tabular}{|c|c|c|c|c|c|c|}
\hline
Rank  & a & b & c & d \\
\hline
1 & 29 & 28 & 43 & 0  \\
2 & 28 & 29 & 0 & 43  \\
3 & 22 & 21 & 57 & 0  \\
4 & 21 & 22 & 0 & 57  \\
\hline
\end{tabular}
\end{table}

Given a preference profile, we determine function $h(a,b)$ as the number of voters that prefer candidate $a$ over candidate $b$,
\begin{equation*}
h(a,b) = | \{ i : a \succ_i b, \quad 1 \leq i \leq n  \} |, \quad
a,b\in A, \quad a\neq b.
\end{equation*}

Matrix $h$ with elements $h(a,b)$ is called a \textbf{tournament
matrix}. (Note that $h(a,b)=n-h(b,a)$ for each $a\neq b$.)

Table~\ref{profile_main} provides the tournament matrix for the profile in Example~\ref{example_main}.

We say that candidate $a$ \textbf{wins in a pairwise comparison} to candidate~$b$, if $h(a,b)>n/2$.

For some subset $B\subseteq A$, a candidate is called a \textbf{Condorcet winner} \citep{Condorcet1785},\footnote{The collection \citep{McLean95} contains English translations of original works by Borda, Condorcet, Nanson, Dodgson, and other early researches.} if he/she wins in a pairwise comparison to each candidate in this subset. Thus, the set of Condorcet winners is
\begin{equation*}
CW(B)=\{b\in B : h(b,a) > n/2 \quad \mbox{for each} \quad a\in
B\setminus b \}, \quad B\subseteq A.
\end{equation*}

It is easy to see that the set of Condorcet winners $CW$ is either a singleton or empty.

For the preference profile in Table~\ref{profile_main}, candidate~$a$ is the Condorcet winner.

Similarly, we say that candidate $a$ \textbf{weakly wins in a pairwise comparison} to candidate~$b$, if $h(a,b) \geq n/2$. For some subset $B\subseteq A$, a candidate is called a \textbf{weak Condorcet winner} if he/she \emph{weakly} wins in a pairwise comparison to each candidate in this subset.

Let a \textbf{positional vector} of candidate~$a$ be vector $n(a) = (n_1(a),\ldots, n_m(a))$, where $n_l(a)$ is the number of voters for whom candidate $a$ has rank $l$ in individual preferences,
\begin{equation*}
n_l(a) = |\{ i : R_i^a = l, \quad 1 \leq i \leq n \}|, \quad a\in A,
\quad l\in \{1,\ldots, m\}.
\end{equation*}

The definition implies that each positional vector has non-negative elements, $n_l(a)\geq 0$ for each $l$, and the sum of elements is equal to the number of voters $\sum\limits_{l=1}^m {n_l(a) = n}$.

Candidate $a$ is called a \textbf{majority winner}, if $n_1(a)>n/2$. Similarly, candidate $a$ is called a \textbf{majority loser} if $n_m(a)>n/2$.

A collection of positional vectors for all candidates is called a
\textbf{positional matrix} $n(\succ) = (n(a_1), \ldots, n(a_m))$.

Table~\ref{profile_main} provides the positional matrix for the profile in Example~\ref{example_main}.

A mapping $C(B,{\succ})$ that to each nonempty subset $B\subseteq A$
and each preference profile~$\succ$ gives a \textbf{choice set} is called a \textbf{voting rule} (or social choice rule),\footnote{Any social choice rule is a voting rule. There exist voting rules that are not social choice rules; for example, approval voting, preference approval voting \citep{Brams09}, and majority judgement \citep{BalinskiLaraki11}.}
\begin{equation*}
C : 2^A \setminus \emptyset \times L(A)^n \rightarrow 2^A,
\end{equation*}
where $C(B,\succ)\subseteq B$ for any $B$; and $C(B,\succ) = C(B,\succ')$, whenever preference profiles $\succ, \succ'$ coincide on $B$.

A rule is called \textbf{universal} if $C(B,\succ)\neq \emptyset$ for any nonempty~$B$ and any profile~$\succ$. For instance, the Condorcet rule $CW(B,\succ)$ is not universal.

Let us define the criteria that are critical for the results of the paper and the voting rules considered below.\footnote{The ``extremely desirable'' criteria of universality, non-imposition, anonymity, neutrality, unanimity are satisfied by all voting rules considered in this paper \citep{Fischer16,Taylor05,Tideman06,Zwicker16}.}

\textbf{Majority criterion}. For each preference profile, if some candidate~$a$ is top-ranked by more than half the voters ($n_1(a)>n/2$) then the choice set coincides with this candidate.

\textbf{Mutual majority criterion}.\footnote{Mutual majority criterion is implied by a more general axiom for multi-winner voting called Droop-Proportionality for Solid Coalitions \citep{Woodall97}.} For each preference profile, if more than half the voters give to some $k$ candidates ($B = \{b_1,\ldots, b_{k}\}$, $1\leq k < m$) top $k$ ranks in an arbitrary order then the choice set is included in~$B$.

\textbf{Majority loser criterion}.\footnote{For a voting rule, the majority criterion is satisfied if and only if the \emph{absolute majority winner paradox} never occurs. For a voting rule, the majority loser criterion is satisfied if and only if the \emph{absolute majority loser paradox} never occurs \citep{FelsenthalNurmi18,Diss18}.} For each preference profile, if some candidate~$a$ is bottom-ranked by more than half the voters ($n_m(a)>n/2$) then the choice set excludes this candidate.

For any fixed quota $q\in (0,1)$ and any fixed number of preferred candidates $k$ among the total of $m$ candidates, we define the next criteria.

\textbf{($q,k,m$)-majority criterion}.\footnote{($q,k,m$)-majority criterion is even more general than the concept q-PSC formalized by \citet{AzizLee17} if the latter is applied to single-winner elections. The weak mutual majority criterion defined by \citet{Kondratev18} is a particular case of $q=k/(k+1)$. Also, q-majority decisiveness proposed by \citet{BaharadNitzan02} is a particular case of $k=1$. A somewhat similar approach but for q-Condorcet consistency is developed by \citet{BaharadNitzan03}, \citet{Courtin15}, and \citet{MahajneVolij18}. All these approaches are based on worst-case analysis.} For each preference profile with a total of $m$ candidates, if a share of voters higher than $q$ gives to some $k$ candidates ($B = \{b_1,\ldots, b_{k}\}$, $1\leq k < m$) top $k$ ranks in an arbitrary order then the choice set is included in~$B$.

For a given $q,k$, we say that a rule satisfies \textbf{($q,k$)-majority criterion} if it satisfies ($q,k,m$)-majority criterion for each~$m$.

For a given $q$, we say that a rule satisfies \textbf{$q$-mutual majority criterion} if it satisfies ($q,k$)-majority criterion for each~$k$.

For universal voting rules, it is also apparent from the definitions that the mutual majority criterion implies the majority criterion; for any $k, m$ and any $q'\geq q$, ($q,k,m$)-majority criterion implies ($q',k,m$)-majority criterion; the majority criterion is equivalent to ($1/2,1$)-majority criterion; the mutual majority criterion is equivalent to $(1/2,k)$-majority criterion with arbitrary~$k$; the majority loser criterion is equivalent to ($1/2,m-1,m$)-majority criterion.\footnote{Also, one can see that the unanimity criterion is equivalent to ($1-\varepsilon,1$)-majority criterion with an infinitely small~$\varepsilon>0$.}

\subsection{Voting Rules}\nopagebreak

This subsection presents the definitions of the voting rules that satisfy the majority criterion but do not satisfy the mutual majority criterion. We also consider monotonic scoring rules that generally do not satisfy the majority criterion.

In the \textbf{plurality voting rule} the candidate that receives the most top positions is declared the winner,
\begin{equation*}
\mathrm{Pl}(A,\succ) = \{a\in A : n_1(a)\geq n_1(b) \quad \mbox{for each} \quad b\in A\setminus a \}.
\end{equation*}

The \textbf{plurality with runoff (RV)} voting rule proceeds in two rounds: first, the two candidates with the most top positions are determined, then, the winner is chosen between the two using the simple majority rule.

According to \textbf{Simpson's rule} (\citeyear{Simpson69}), aka maximin voting rule \citep[see also][]{Young77}, each candidate receives a score
equal to the minimal number of votes that this candidate gets
compared to any other candidate,
\begin{equation*}\label{Si}
Si(a) = \min_{b\in A\setminus \{ a \}} {h(a,b)},
\end{equation*}
and the winner is the candidate with the highest score.

The \textbf{Young} score of candidate $a$ is defined as the smallest number $n'$ such that there is a set of $n'$ voters such that $a$ is a \emph{weak} Condorcet winner when those $n'$ voters are removed from the election \citep{Young77,Caragiannis16}. All candidates with the lowest Young score in a given election are its Young winner(s).

According to the \textbf{Condorcet least-reversal rule (CLR)} \citep[aka the simplified Dodgson rule, see][]{Tideman06} the winner is, informally, the candidate $a\in A$ that needs the least number of reversals in pairwise comparisons in order to become a Condorcet winner. Formally, the winner $d$ minimizes the following sum of losing margins compared to each candidate $c$:
\begin{equation}\label{Clr}
p^{CLR}_d=\sum\limits_{c\in A\setminus d}{\max \left\{\frac{n}{2} - h(d,c), 0\right\}}.
\end{equation}

The \textbf{Dodgson} score of candidate $a$ is the smallest number of sequential exchanges of adjacent candidates in preference orders such that after those exchanges $a$ is a Condorcet winner \citep{Dodgson1876,McLean95,Caragiannis16}. All candidates with the smallest Dodgson score are the Dodgson winner(s).

The \textbf{proportional veto core} is defined as follows \citep{Moulin81,Moulin83book}. For a given profile with $n$ voters and $m$ candidates, a candidate~$a$ is \emph{not stable} if some coalition of $t$~voters blocks him. Blocking means that there is a subset $B$ of candidates such that each of the $t$ voters prefers each candidate $b\in B$ over candidate $a$ and $B$ is large enough, $|A\setminus B| < m\cdot t/n$.

The proportional veto core consists of all \emph{stable} candidates.

Under the \textbf{Borda rule} \citep{Borda1781,McLean95}, the first-best candidate in an individual preference gets $m-1$ points, the second-best candidate gets $m-2$, \ldots, the last gets $0$ points.

The total Borda score can be calculated using the positional vector $n(a)$
as follows:
\begin{equation}\label{Bo_pos}
Bo(a)=\sum\limits_{i=1}^m{n_i(a)(m-i)}, \quad a\in A.
\end{equation}

The candidate with the highest total score wins. The score can
also be calculated using the tournament matrix:
\begin{equation}\label{Bo_TM}
Bo(a)= \sum_{b\in A\setminus \{ a \}} {h(a,b)}, \quad a\in A.
\end{equation}

\textbf{Black's rule} (\citeyear{Black58}) selects a Condorcet winner. If a Condorcet winner does not exist then the candidate with the highest Borda score~(\ref{Bo_TM}) is selected.

In a \textbf{non-generalized scoring rule} each of $m$ candidates is assigned a score from $s_1, \ldots , s_m$ for a corresponding position in a voter's individual preference and then the scores are summed up over all voters.\footnote{In each \emph{generalized scoring rule}, tie-breaking is performed using a sequence of non-generalized scoring rules \citep{Smith73,Young75}.} In the paper, we consider \textbf{monotonic scoring rules} in which $s_1 > s_m$ and $s_1 \geq s_2 \geq \ldots \geq s_m$. The plurality rule and the Borda rule are monotonic scoring rules with the scores $s_1=1, s_2 = \ldots = s_{m} = 0$ and $s_1 = m-1, s_2 = m-2, \ldots s_m = 0$, respectively.

\textbf{Inverse plurality rule} (aka \textbf{anti-plurality rule} or \textbf{negative voting}) is a monotonic scoring rule with the scores $s_1 = \ldots = s_{m-1} = 1, s_m =0$. 

If the difference in scores is positive and nondecreasing from position $m$ to position~1, that is, $0<s_{m-1}-s_m \leq s_{m-2}-s_{m-1}\leq \ldots \leq s_1-s_2$, then a rule will be called a \textbf{convex scoring rule}. For instance, the Borda rule is a convex scoring rule.

Our last rule is based on \emph{truncated Borda scores} defined as follows. For some positional vector $n(a)$ and some real number $t\in (0,+\infty)$ the truncated Borda score \citep{Fishburn74} 
\begin{equation*}\label{B_t}
B_t(a) = t \cdot n_1(a) {+} (t{-}1) n_2(a) {+} \ldots {+} (t {-} \lfloor{t}\rfloor) n_{\lfloor{t}\rfloor+1}(a), \quad t\in (0{,}{+}\infty),
\end{equation*}
where formally put $n_i(a)=0$ for $i>m$. The definition implies that
$B_{m-1}(a)=Bo(a)$.

Now we can define a modification of a standard median voting rule \citep{Sertel99}: the \textbf{convex median voting rule (CM)} \citep{Kondratev18}. Instead of the standard sequence $n_1(a),n_1(a) + n_2(a),n_1(a) + n_2(a) + n_3(a), \ldots$ we use the truncated Borda scores. If $n_1(a)>n/2$ for some candidate~$a$ then this candidate is the winner. Otherwise, for each candidate~$a$ define the score of the convex median using the following formula:
\begin{equation*}
    \mathrm{CM}(a) = \max\left\{ t\geq1: \frac{B_t(a)}{t} \leq \frac{n}{2} \right\},
\end{equation*}
and the winner is the candidate with the lowest value of the convex median.

We complete this section with one more desirable criterion.

\textbf{Second-order positional dominance (2-PD)} \citep{Stein94}. Whenever candidate~$a$ obtains a higher score than candidate~$b$ for all convex scoring rules then candidate~$b$ is not included in the choice set.

\section{Results}\nopagebreak
\label{Main}

\subsection{Majority-Consistent Voting Rules}

This subsection considers the voting rules that satisfy the majority criterion (thus, they satisfy ($q,k$)-majority criterion with $k=1$ and any $q\geq 1/2$) but do not satisfy the mutual majority criterion.\footnote{For completeness of results, we should mention well-studied voting rules that satisfy the mutual majority criterion. These are the Condorcet extensions: Nanson's (\citeyear{Nanson1882}), see also \citet{McLean95}, Baldwin's (\citeyear{Baldwin1926}), maximal likelihood \citep{Kemeny59}, ranked pairs \citep{Tideman87}, Schulze's (\citeyear{Schulze11}), successive elimination \citep[see e.g.,][]{FelsenthalNurmi18}, and those tournament solutions which are refinements of the top cycle \citep{Good71,Schwartz72}; other rules include the single transferable vote \citep{Hare1859}, Coombs' (\citeyear{Coombs64}), Bucklin's \citep[see e.g.,][]{FelsenthalNurmi18}, median voting rule \citep{Bassett99}, majoritarian compromise \citep{Sertel99}, q-approval fallback bargaining \citep{Brams01}. For their formal definitions and properties, we also advise \citet{Brandt16}, \citet{FelsenthalNurmi18}, \citet{Fischer16}, \citet{Taylor05}, \citet{Tideman06}, and \citet{Zwicker16}.\label{footnote-MM-rules}}
In the case of only two candidates, each rule satisfying the majority criterion coincides with the \textbf{simple majority rule} where the winner is the candidate that gets at least half the votes.\footnote{In case of only $m=2$ candidates the simple majority rule is the most natural as it satisfies a number of other important axioms according to May's Theorem (\citeyear{May52}).} In what follows we consider the case of $m>2$ candidates.

For the voting rules, below we find necessary and sufficient conditions under which the $(q,k,m)$-majority criterion is satisfied. First we determine the tight bounds for the plurality rule, Simpson's rule, Young's rule and the Condorcet least-reversal rule.

\begin{theorem}\label{theorem_plurality}
For each $m > k\geq 1$, the plurality rule satisfies ($q,k,m$)-majority criterion if and only if $q\geq k/(k+1)$.
\end{theorem}

\begin{theorem}\label{theorem_Simpson}
For each $m>k>1$, Simpson's rule satisfies ($q,k,m$)-majority criterion if and only if $q\geq (k-1)/k$.\footnote{This tight bound $q=(k-1)/k$ for Simpson's rule coincides with the tight bound of the q-majority equilibrium \citep{Greenberg79,Kramer77}, and with the minimal quota, that guarantees the acyclicity of preferences \citep{Craven71,Ferejohn74,Usiskin64}.}
\end{theorem}

\begin{theorem}\label{theorem_Young}
For each $m>k>1$, Young's rule satisfies ($q,k,m$)-majority criterion if and only if $q\geq (k-1)/k$.
\end{theorem}

\begin{theorem}\label{theorem_CLR}
For each $m>k\geq 2$ and for each even $k$, the Condorcet least-reversal rule satisfies ($q,k,m$)-majority criterion if and only if $q\geq (5k-2)/(8k)$; for each $m>k\geq 1$ and for each odd $k$, the rule satisfies the criterion if and only if $q\geq (5k^2-2k+1)/(8k^2)$.
\end{theorem}

A few peculiarities can be observed regarding the results above. For the plurality rule, Simpson's rule, Young's rule, and the Condorcet least-reversal rule, the minimal size of the qualified mutual majority $q(k,m)$ depends on the number of preferred candidates $k$ but does not depend on the total number of candidates~$m$. Theorem~\ref{theorem_CLR} shows that the Condorcet least-reversal rule is, perhaps surprisingly, very close to satisfying the mutual majority criterion as it satisfies $5/8$-mutual majority criterion. At the other extreme, the plurality rule gives the highest non-trivial minimal quota for a given $k>1$ among all studied voting rules.

Interestingly, both Simpson's rule and Young's rule have the same minimal quota $q$ for each number of preferred candidates $k$ and each total number of candidates $m$. This is the only such coincidence among all the rules that we consider.

Next, we determine the tight bounds for the plurality with runoff and Black's rule.

\begin{theorem}\label{theorem_RV}
For each $m-1 = k \geq 1$, the plurality with runoff rule satisfies ($q,k,m$)-majority criterion if and only if $q\geq 1/2$; for each $m-1 > k >1$, the rule satisfies the criterion if and only if $q \geq k/(k+2)$.
\end{theorem}

\begin{theorem}\label{theorem_Black}
For each $m>k>1$, Black's rule satisfies ($q,k,m$)-majority criterion if and only if $q\geq (2m-k-1)/(2m)$.
\end{theorem}

Theorem~\ref{theorem_Black} actually finds the tight bound of quota for the \textbf{Borda rule}. In particular, in case $k=1$, this quota equals $q=(m-1)/m$, and was also calculated by \citet{BaharadNitzan02}, \citet{Nitzan09}.

\begin{theorem}\label{theorem_convex_median}
For each $m>2k$, the convex median voting rule satisfies ($q,k,m$)-majority criterion if and only if $q\geq (3k-1)/(4k)$; for each $m=k+1$ -- if and only if $q\geq 1/2$; for each $2k\geq m > k+1$, the tight bound~$q$ satisfies the inequality $\frac{1}{2} < q < \frac{3k-1}{4k}$ and also the equation 
\begin{equation}\label{ThCM}
4k(m-k-1)q^2 + (5k^2+5k-2mk-m^2+m)q + m(m-1-2k) = 0.
\end{equation}
\end{theorem}

Let us briefly motivate the results for the convex median voting rule. The Borda rule satisfies second-order positional dominance (2-PD) but it fails the majority criterion. The convex median voting rule was proposed by \citet{Kondratev18} as a rule that satisfies both 2-PD and the majority criterion. Theorem~\ref{theorem_convex_median} shows that this rule is much closer to satisfying the stronger criterion of the mutual majority as it satisfies $3/4$-mutual majority criterion.

For Dodgson's rule, below we find sufficient conditions. Necessary and sufficient conditions remain an open question.

\begin{theorem}\label{theorem_Dodgson}
For $m>k\geq 1$, Dodgson's rule satisfies ($q,k,m$)-majority criterion with $q\geq k/(k+1)$; the rule fails the criterion with $q<(5k-2)/(8k)$ in the case of evens $k\geq 2$, and $ q <(5k^2-2k+1)/(8k^2)$ in the case of odds $k\geq 1$.
\end{theorem}

\begin{table}
\caption{Measuring majority power: for a given $k$, the minimal quota $q$ such that ($q,k$)-majority criterion is satisfied}
\label{table-k-1-2-3-4-5}
\begin{tabular}{|c|c|c|c|c|c|c|}
\hline
Voting rule & $k=1$ & $k=2$ & $k=3$ & $k=4$ & $k>1$ & $\sup_{k} q$ \\
\hline
Instant-runoff & 0.500 & 0.500 & 0.500 & 0.500 & 0.500 & 0.500 \\
CLR (even $k$)  &  & 0.500 &  & 0.563 & $(5k-2)/(8k)$ & 0.625 \\
CLR (odd $k$)  & 0.500 &  & 0.556 & & $(5k^2-2k+1)/(8k^2)$ & 0.625 \\
Convex median & 0.500 & 0.625 & 0.667 & 0.688 & $(3k-1)/(4k)$ & 0.750 \\
RV  & 0.500 & 0.500 & 0.600 & 0.667 & $k/(k+2)$ & 1.000 \\
Simpson's & 0.500 & 0.500 & 0.667 & 0.750 & $(k-1)/k$ & 1.000 \\
Young's & 0.500 & 0.500 & 0.667 & 0.750 & $(k-1)/k$ & 1.000 \\
Plurality & 0.500 & 0.667 & 0.750 & 0.800 & $k/(k+1)$ & 1.000 \\
Black's  & 0.500 & 1.000 & 1.000 & 1.000 & 1.000 & 1.000 \\
Proportional veto  & 1.000 & 1.000 & 1.000 & 1.000 & 1.000 & 1.000 \\
Borda  & 1.000 & 1.000 & 1.000 & 1.000 & 1.000 & 1.000 \\
Inverse plurality  & 1.000 & 1.000 & 1.000 & 1.000 & 1.000 & 1.000 \\
\hline
\end{tabular}

\vspace{0.2cm}

\justify
\footnotesize{\emph{Notes}: The following notations are used: CLR -- Condorcet least-reversal rule, RV -- plurality with runoff rule. The voting rules are ordered according to the minimal size of the qualified mutual majority $q$ for $k>4$. For majority-consistent rules $q=1/2$ whenever $k=1$. The instant-runoff voting satisfies the mutual majority criterion and therefore $q=1/2$ for each~$k$.}

\end{table}

The summary of the results from this subsection (Theorem \ref{theorem_plurality} to \ref{theorem_convex_median}) is presented in Fig.~\ref{partial-order}. We can only \emph{partially} order the voting rules with respect to majority power when the total number of candidates $m$ and the number of preferred candidates $k$ are not specified. The minimal size of the qualified mutual majority $q(k,m)$ weakly increases along the arrows.

Using the $(q,k)$-majority criterion we get a \emph{complete} order of the voting rules, as is shown in Table~\ref{table-k-1-2-3-4-5}. The voting rules are ordered based on the quota $q$ from those with the highest majority power to those with the lowest whenever the number of preferred candidates $k>4$. When $k=1$, the quota for majority-consistent rules equals 0.5. (The proportional veto core, the Borda rule and the inverse plurality rule do not satisfy the majority criterion and have quota~1.) As $k$ increases and the mutual majority's preferences over the preferred candidates might become more diverse, the minimal quota $q$ also weakly increases. The rate of this increase varies among the rules and for a small $k\in\{2,3,4\}$ the order of the rules varies as well.

\subsection{Majority-Inconsistent Voting Rules.}

In this subsection we present the results of the proportional veto core and scoring voting rules. 

\begin{theorem}\label{theorem_PVC}
For each $m > k\geq 1$, the proportional veto core satisfies ($q,k,m$)-majority criterion if and only if $q\geq (m-k)/m$.
\end{theorem}

The result above is not surprising because forcing the selection of one of the $k$ most preferred candidates is equivalent to the vetoing of $m-k$ least preferred candidates. Hence, any voting rule which selects from the proportional veto core also satisfies Theorem~\ref{theorem_PVC}.\footnote{Using the same arguments as in the proof of Theorem~\ref{theorem_PVC}, one can check that voting rules implemented by sequential voting by veto \citep{Mueller78,Moulin82,Moulin83book,FelsenthalMachover92} have the same tight bound on the size of the qualified mutual majority~$q$.}  

Also, this subsection generalizes Theorem~\ref{theorem_plurality} for the plurality rule ($s_1=1, s_2=\ldots = s_m = 0$) and Theorem~\ref{theorem_Black} for the Borda rule ($s_i=m-i, \; i=1,\ldots,m$). 

\begin{theorem}\label{theorem_scoring}
For each $m>k\geq 1$, a monotonic scoring rule satisfies ($q,k,m$)-majority criterion if and only if the quota $q$ satisfies the next inequality
\begin{equation}\label{qMSR}
q \geq \frac{s_1 - \frac{1}{k} \sum\limits_{i=1}^k{s_{m-i+1}}}{s_1-
\frac{1}{k} \sum\limits_{i=1}^k{s_{m-i+1}} + \frac{1}{k}
\sum\limits_{i=1}^k{s_i} - s_{k+1}}.
\end{equation}

For each $m>1$, a monotonic scoring rule satisfies the majority loser criterion\footnote{Equivalently, the \emph{absolute majority loser paradox} never occurs.} if and only if the next inequality holds
\begin{equation}\label{ML-inequality}
s_1-\frac{s_2+\ldots+s_m}{m-1}\leq \frac{s_1+\ldots+s_{m-1}}{m-1}-s_m.
\end{equation}
\end{theorem}

In particular, in case $k=1$, the inequality (\ref{qMSR}) is\footnote{Equivalently, $q$-majority consistency (aka $q$-majority decisiveness) is satisfied.}
\begin{equation*}
  q \geq \frac{s_1-s_m}{s_1-s_m+s_1-s_2},
\end{equation*}
and was also calculated by \citet{BaharadNitzan02}, \citet{Nitzan09}.

In particular, in case $k=1$ and $q=1/2$, we receive the inequality $s_m\geq s_2$. Thus, for each $m>1$, a monotonic scoring rule satisfies the majority criterion if and only if this rule is equivalent to the plurality rule ($s_1>s_2=\ldots= s_m$). This fact was also established by \citet{Lepelley92}, \citet{Sanver02}.

In particular, in case $m=3$, the inequality (\ref{ML-inequality}) was established, for example, by \citet{Diss18}.

For the inverse plurality rule ($s_1=\ldots=s_{m-1}=1, s_m=0$), Theorem~\ref{theorem_scoring} directly implies the next statement.

\begin{theorem}\label{theorem_inverse_plurality}
For each $m-1>k\geq 1$ and for each $q<1$, the inverse plurality rule fails ($q,k,m$)-majority criterion; for each $m-1=k\geq 1$, the rule satisfies ($q,k,m$)-majority criterion if and only if $q\geq 1/m$.
\end{theorem}

Note that for the inverse plurality rule and the proportional veto core the minimal quota~$q$ may be below one half. For the inverse plurality rule, this occurs whenever the number of preferred candidates $k$ equals the total number of candidates $m$ minus one, which is the same as saying that one candidate is vetoed by the mutual majority. We present the analysis of the veto power in the next subsection.

\begin{table}
\caption{Minimal quota $q$ such that ($q,k,m$)-majority criterion is satisfied}
\label{table-m-3-4}
\begin{tabular}{|c|c|c|c|c|c|}
\hline
& $m=3$ & $m=3$ & $m=4$ & $m=4$ & $m=4$ \\
Voting rule & $k=1$ & $k=2$ & $k=1$ & $k=2$ & $k=3$ \\
\hline
Instant-runoff & 0.500 & 0.500 & 0.500 & 0.500 & 0.500 \\
Condorcet least-reversal  & 0.500 & 0.500 & 0.500 & 0.500 & {\bf 0.556} \\
Convex median & 0.500 & 0.500 & 0.500 & {\bf 0.593} & 0.500 \\
Plurality with runoff  & 0.500 & 0.500 & 0.500 & 0.500 & 0.500 \\
Simpson's & 0.500 & 0.500 & 0.500 & 0.500 & {\bf 0.667} \\
Young's & 0.500 & 0.500 & 0.500 & 0.500 & {\bf 0.667} \\
Plurality & 0.500 & {\bf 0.667} & 0.500 & {\bf 0.667} & {\bf 0.750} \\
Black's  & 0.500 & 0.500 & 0.500 & {\bf 0.625} & 0.500 \\
Proportional veto & {\bf 0.667} & {\bf 0.333} & {\bf 0.750} &  0.500 & {\bf 0.250} \\
Borda  & {\bf 0.667} & 0.500 & {\bf 0.750} & {\bf 0.625} & 0.500 \\
Inverse plurality & {\bf 1.000} & {\bf 0.333} & {\bf 1.000} & {\bf 1.000} & {\bf 0.250} \\
\hline
\end{tabular}
\vspace{0.2cm}

\justify
\footnotesize{\emph{Notes}: The voting rules are ordered as in Table~\ref{table-k-1-2-3-4-5}. We highlight the instances of quotas that are different from 0.5.}

\end{table}

Theorems \ref{theorem_PVC} and \ref{theorem_scoring} extend the class of voting rules that we can compare using the minimal quota $q$, yet this comparison should be made for a specific number of candidates $m$ (as for most monotonic scoring rules $(q,k)$-majority criterion gives the same tight quota $q=1$). Thus, we have to use the ($q,k,m$)-majority criterion. Table~\ref{table-m-3-4} presents this comparison for the majority-consistent voting rules that we considered earlier, the proportional veto core, the Borda rule, and the inverse plurality rule when the total number of candidates $m=3,4$. 

We see that for majority-consistent voting rules (except the plurality rule), when $m,k$ are small, most of the values of the minimal quota $q$ is equal to 0.5. Interestingly, whenever $m\leq 4$ (Table~\ref{table-m-3-4}) or $k\leq 2$ (Table~\ref{table-k-1-2-3-4-5}), the plurality with runoff rule has exactly the same minimal quotas $q=0.5$ as the instant-runoff voting. 

We can illustrate the results for the Borda rule and the plurality rule using Example~\ref{example_main} presented in Table~\ref{profile_main}. Candidates $a$ and $b$ are the two ($k=2$) preferred candidates among the total of four ($m=4$) candidates, supported by the mutual majority of 57\% voters. This value is below the minimal quota for the Borda rule ($q=0.625$) and the plurality rule ($q=0.667$), and thus these rules might not select $a$ or $b$. In our example both rules select candidate~$c$.

\subsection{Veto Power}

The criteria that we presented above for majority power also allow us to state a somewhat opposite research question, that is, of veto power. Specifically: how large should a group of voters be in order to be able to block its $l$ least preferred candidates? 

This problem is dual to the problem of finding the minimal quota for the mutual majority that has $k=m-l$ preferred candidates. Thus, we can immediately compute the minimal quota of such a group as in the $(q,m-l,m)$-majority criterion.\footnote{Previously, the concept of veto power in voting was introduced by \citet{BaharadNitzan05,BaharadNitzan07b} for settings with $l=1$ and by \citet{Moulin81,Moulin82,Moulin83book} for settings with an arbitrary~$l$. Their concepts are also based on the worst-case analysis but are different from ours in that they involve strategic voting.}

Let us define the veto criterion formally. For a given quota $q$ and a given number of the least preferred candidates~$l$, we say that a rule satisfies the \textbf{($q,l$)-veto criterion} if it satisfies the ($q,m-l,m$)-majority criterion for each~$m$.

Overall, when we compare the rules based on the veto criterion from the most veto-preserving to the least, we get a \emph{partial} order (the same as for $(q,k,m)$-majority criterion, see Fig.~\ref{partial-order}). However, when we compare the rules based on $(q,l)$-veto criterion, that is, for an arbitrary total number of candidates $m$, we get a \emph{complete} order for $l>2$ as shown in Table~\ref{table-veto}. In case $l=1$ this order is different and is very peculiar, as we discuss below.

\begin{table}
\caption{Measuring veto power: for a given $l$, the minimal quota $q$ such that the ($q,l$)-veto criterion is satisfied}
\label{table-veto}
\begin{tabular}{|c|c|c|c|c|c|c|}
\hline
Voting rule & $l=1$ & $l=2$ & $l=3$ & $l=4$ & $l>3$ & $\sup_{l} q$ \\
\hline
Instant-runoff & 0.500 & 0.500 & 0.500 & 0.500 & 0.500 & 0.500 \\
Condorcet least reversal  & 0.625 & 0.625 & 0.625 & 0.625 & 0.625 & 0.625 \\
Convex median & 0.500 & 0.593 & 0.640 & 0.667 & $(3l-4)/(4l-4)$ & 0.750 \\
Black's  & 0.500 & 0.625 & 0.700 & 0.750 & $(2l+1)/(2l+4)$ & 1.000 \\
Proportional veto  & 0.333 & 0.667 & 0.750 & 0.800 & $l/(l+1)$ & 1.000 \\
Borda  & 0.500 & 0.667 & 0.750 & 0.800 & $l/(l+1)$ & 1.000 \\
Inverse plurality & 0.333 & 1.000 & 1.000 & 1.000 & 1.000 & 1.000 \\
Plurality with runoff  & 0.500 & 1.000 & 1.000 & 1.000 & 1.000 & 1.000 \\
Simpson's & 1.000 & 1.000 & 1.000 & 1.000 & 1.000 & 1.000 \\
Young's & 1.000 & 1.000 & 1.000 & 1.000 & 1.000 & 1.000 \\
Plurality & 1.000 & 1.000 & 1.000 & 1.000 & 1.000 & 1.000 \\
\hline
\end{tabular}

\vspace{0.2cm}

\justify
\footnotesize{\emph{Notes}: The total number of candidates $m\geq 3$. The voting rules are ordered according to the minimal size of the qualified mutual majority $q$ for $l>2$ least preferred candidates. For majority loser-consistent rules $q \leq 1/2$ whenever $l=1$. The instant-runoff voting satisfies the mutual majority criterion and therefore $q=1/2$ for each~$l$.}
\end{table}

When $l=1$, our results highlight the inverse plurality rule as the rule that respects the minorities the most: its minimal quota is $q=1/m$. The previous literature arrived at the same conclusion by comparing only the monotonic scoring rules \citep{BaharadNitzan05,BaharadNitzan07a,BaharadNitzan07b}. We extend the comparison to non-scoring rules and confirm this consensus when $l=1$. 
 
However, when $l>1$, we arrive at the opposite conclusion. In this case, the minimal quota $q$ for inverse plurality rule jumps to 1 and thus no group of voters (except the entire set) can veto $l>1$ candidates (see Table~\ref{table-m-3-4} where $k=m-l$, Table~\ref{table-veto}, and Theorem~\ref{theorem_inverse_plurality}). The reason is that unless the group coordinates, some of the $l$ candidates may receive a very small number of lowest positions.

Comparing the results for the veto criterion in Table~\ref{table-veto} with the quotas for the $(q,k)$-majority criterion in Table~\ref{table-k-1-2-3-4-5} we see that voting rules differ in their power for the most preferred candidates and the least preferred candidates. Overall, the order of rules remains the same except for the proportional veto core, Black's rule, the Borda rule, and the inverse plurality rule. These rules perform better when a group of voters need to veto a candidate, rather than make him win. 

Surprisingly, the proportional veto core does not give the highest veto power according to the $(q,l)$-veto criterion. This is because for each given number of the least preferred candidates $l$ the worst case arises when the total number of candidates $m=l+1$. In this case voters need to veto all but one candidate, which is the same as forcing the win of the remaining candidate.

The situation changes if we restrict our analysis to the case $l\leq m/2$ (see Table~\ref{table-veto-2}), that is, when one can veto only her bottom half of the candidates. Comparing the results in Table~\ref{table-veto} and Table~\ref{table-veto-2} we see that only the proportional veto core changes its relative position and has the highest opportunities to veto not more than half the candidates.

\begin{table}
\caption{Measuring veto power: for a given $l$, the minimal quota $q$ such that the ($q,m-l,m$)-majority criterion is satisfied whenever $l\leq m/2$ and $m\geq 3$}
\label{table-veto-2}
\begin{tabular}{|c|c|c|c|c|c|c|}
\hline
Voting rule & $l=1$ & $l=2$ & $l=3$ & $l=4$ & $l>3$ & $\sup_{l} q$ \\
\hline
Proportional veto & 0.333 & 0.500 & 0.500 & 0.500 & 0.500 & 0.500 \\
Instant-runoff & 0.500 & 0.500 & 0.500 & 0.500 & 0.500 & 0.500 \\
Condorcet least-reversal  & 0.625 & 0.625 & 0.625 & 0.625 & 0.625 & 0.625 \\
Convex median & 0.500 & 0.593 & 0.640 & 0.667 & $\frac{-7+3l+\sqrt{17-10l+9l^2}}{8l-8}$ & 0.750 \\
Black's  & 0.500 & 0.625 & 0.667 & 0.688 & $(3l-1)/(4l)$ & 0.750 \\
Borda  & 0.500 & 0.625 & 0.667 & 0.688 & $(3l-1)/(4l)$ & 0.750 \\
Inverse plurality & 0.333 & 1.000 & 1.000 & 1.000 & 1.000 & 1.000 \\
Plurality with runoff  & 0.500 & 1.000 & 1.000 & 1.000 & 1.000 & 1.000 \\
Simpson's & 1.000 & 1.000 & 1.000 & 1.000 & 1.000 & 1.000 \\
Young's & 1.000 & 1.000 & 1.000 & 1.000 & 1.000 & 1.000 \\
Plurality & 1.000 & 1.000 & 1.000 & 1.000 & 1.000 & 1.000 \\
\hline
\end{tabular}

\vspace{0.2cm}

\justify
\footnotesize{\emph{Notes}: The total number of candidates $m\geq 3$. The voting rules are ordered according to the minimal size of the qualified mutual majority $q$ for $l>2$ least preferred candidates. For majority loser-consistent rules $q \leq 1/2$ whenever $l=1$. The instant-runoff voting satisfies the mutual majority criterion and therefore $q=1/2$ for each~$l$. Under the proportional veto core any simple majority can veto half the candidates and therefore $q=1/2$ for each~$l>1$.}
\end{table}

\subsection{Other Criteria and Tradeoffs}\label{ssec_tradeoffs}

In this subsection, we discuss the tradeoffs between the $(q,k)$-majority criterion, the $(q,l)$-veto criterion, and other criteria from the literature. 

For the general classes of voting rules, satisfying the mutual majority criterion (aka $(1/2,k)$-majority criterion with arbitrary~$k$) is not a concern. For instance, among Condorcet-consistent rules we can highlight the ranked pairs rule introduced by \citet{Tideman87} and Schulze's rule (\citeyear{Schulze11}), among iterated positional rules -- the instant-runoff voting, among positional rules -- the median voting rule \citep{Bassett99}, Bucklin's method \citep[see e.g.,][]{Tideman06}, and the majoritarian compromise \citep{Sertel99}. 

In contrast, for monotonic scoring rules the $(q,k)$-majority criterion is often out of reach. Each scoring rule does not satisfy ($1/2,k$)-majority criterion for some~$k$. However, for each fixed $k$, the scoring rule with the scores $s_1=\ldots=s_k=1,s_{k+1}=\ldots=s_m=0$ satisfies ($1/2,k$)-majority criterion. Below, we consider the tradeoffs for scoring rules in more detail.

For a given voting rule, we can see the tradeoff between majority power (including the majority criterion, and the mutual majority criterion), veto power (including the majority loser criterion), and other criteria from its axiomatic characterizations.

\citet{BaharadNitzan05} prove that the inverse plurality rule is the only non-generalized scoring rule that satisfies the minimal veto criterion.\footnote{The fundamental characterization of the class of generalized and non-generalized scoring rules was introduced by \citet{Smith73} and \citet{Young75}. They use the criteria of universality, anonimity, neutrality, and consistency (aka reinforcement) for the generalized scoring rules, and additionally the continuity (aka Archimedean) criterion for the non-generalized scoring rules. Characterizations of specific scoring rules usually involve the fundamental result above; see \citet{ChebotarevShamis98} for a review and \citet{Richelson78} and \citet{Ching96} for the case of the plurality rule.} Though their minimal veto criterion involves strategic candidates and is different from ours, it shows that the quota $q=1/m$ (for the case of $l=1$ least preferred candidates) characterizes the inverse plurality rule. This implies that the $(q,k)$-majority criterion is never satisfied (for each fixed $k$, the minimal quota $q$ for the inverse plurality rule equals 1).

The plurality rule is the only non-generalized scoring rule that satisfies the majority criterion \citep{Lepelley92,Sanver02}. This implies that if a non-generalized scoring rule satisfies the $(1/2,1)$-majority criterion for $k=1$ then for any given $k$ the minimal quota is necessarily $q=k/(k+1)$.

\citet{Sanver02} and \citet{Woeginger03} prove that a generalized scoring rule cannot simultaneously satisfy the majority criterion and the majority loser criterion. This tradeoff can be illustrated by the plurality rule, the Borda rule, and the inverse plurality rule. While the plurality rule satisfies the majority criterion, it fails the majority loser criterion.  In contrast, the Borda rule and the inverse plurality rule satisfy the majority loser criterion and hence fail the majority criterion. 

Other impossibility results involving the majority criterion and the majority loser criterion can be found in Theorem 4.2 in \citet{Kondratev18}. Among 37 different criteria only the second-order positional dominance (2-PD) is resistant to the $1/2$-mutual majority criterion. Specifically, there is no rule that satisfies both the 2-PD and $1/2$-mutual majority criteria. 

This impossibility is easy to see from the preference profile in Table~\ref{profile_main}. Here, candidates $a$ and $b$ are supported by a mutual majority of 57\% of voters. However, candidate~$c$ obtains a higher score than candidates~$a, b$ for all convex scoring rules, i.e., $c$ second-order positionally dominates $a,b$. 

We can generalize the latter result from the $1/2$-mutual majority criterion to the $(q,k)$-majority criterion for any given $k$. The next theorem establishes the tradeoff between 2-PD and ($q,k$)-majority criterion.

\begin{theorem}\label{theorem-tradeoff}
1) If $k \geq 1$ and $q<2k/(3k+1)$ then there is no rule that satisfies the second-order positional dominance and ($q,k$)-majority criteria;\\  2) There exists a rule that satisfies both criteria for each $k \geq 1$ and each $q\geq 2k/(3k+1)$.\footnote{The rule constructed in the proof of Theorem~\ref{theorem-tradeoff} does not satisfy the criteria of the Condorcet loser, majority loser, and reversal symmetry. In contrast, the convex median voting rule satisfies these three criteria  \citep{Kondratev18} but has a higher tight bound of the size of qualified mutual majority according to Theorem~\ref{theorem_convex_median}.} 
\end{theorem}

This theorem also shows that the 2-PD and $q$-mutual majority criteria are compatible if and only if $q\geq 2/3$.

\section{Concluding Remarks}\nopagebreak
\label{Conclusions}

We introduced and studied the quantitative properties of voting rules which we call the $(q,k)$-majority criterion and the $(q,l)$-veto criterion. These criteria allow us to study how decisive each voting rule is, that is, to what extent a voting rule respects majority power and/or veto power.

Our criteria form a partial order over the studied voting rules (see Fig.~\ref{partial-order}). In general, the rule with a higher majority power also has a higher veto power. The \emph{instant-runoff} voting has the highest majority power and veto power, while the \emph{plurality rule} appears to have a relatively low level of majority power and veto power (see tables~\ref{table-k-1-2-3-4-5}, \ref{table-veto}, \ref{table-veto-2}). 

Our results give several new insights and raise a number of open questions; we briefly list them below.


\medskip

\textbf{Minority protection.} Somewhat surprisingly, the \emph{inverse plurality} rule (aka the \emph{anti-plurality} rule or \emph{negative voting}) also gives low veto power while it was previously assumed to give minorities the highest veto power \citep{BaharadNitzan05,BaharadNitzan07a,BaharadNitzan07b}. This discrepancy comes from the fact that veto power was previously assumed to be used against only one least preferred candidate, $l=1$. In this case our results agree with the literature, while when $l>1$ we show that the inverse plurality rule may require the entire set of voters, $q=1$, to veto these $l$ candidates (see Theorem~\ref{theorem_inverse_plurality} and tables~\ref{table-veto}, \ref{table-veto-2}). An open question here is whether this peculiarity of the inverse plurality rule also occurs in the strategic framework of \cite{BaharadNitzan02,BaharadNitzan07b} and \citet{Moulin81,Moulin82,Moulin83book}.

For each specific setting (fixed number of preferred candidates and the total number of candidates) the partial order becomes complete. Surprisingly, in the complete orders the direct relation between the majority power and the veto power disappears. Specifically, the \emph{proportional veto core}, the \emph{Borda rule}, and  \emph{Black's rule} have lower majority power than the plurality rule (see Table~\ref{table-k-1-2-3-4-5}), but at the same time they have a higher veto power than the plurality with runoff (see tables~\ref{table-veto}, \ref{table-veto-2}). This is possible since the three rules are not comparable with the plurality rule and the plurality with runoff in the partial order (see Fig.~\ref{partial-order}). Thus, the proportional veto core, the Borda rule, and Black's rule are, perhaps, the best rules to protect minorities.

Among the three rules, \emph{Black's rule} seems to have a more distinguished set of properties. From a theoretical point of view, the first stage of Black's rule, the Condorcet rule (aka pairwise majority rule), works well because it is strategy-proof in a large domain \citep{CampbellKelly03} and satisfies the independence of irrelevant alternatives \citep{DasguptaMaskin08,Miller18}. Statistically, Black's rule is much less manipulable than the Borda rule \citep{AleskerovKurbanov99,Aleskerov12,Green-Armytage16}, however, the Borda rule provides a slightly higher social utility efficiency \citep{Merrill84}. In three-candidate elections, Black’s rule encounters voting paradoxes with lower frequencies \citep{PlassmannTideman14SCW} and selects the ``best'' candidate with higher frequencies \citep{PlassmannTideman14PC}. In real elections, however, both rules mostly select the same candidate despite the fact that a Condorcet winner usually exists \citep{FeldGrofman92}.

The proportional veto core consistently gives the highest veto power to minorities whenever they have a mutual dislike of no more than half the candidates (see Theorem~\ref{theorem_PVC} and Table~\ref{table-veto-2}). However the practical use of this rule is limited since it is not easy to compute even for a small total number of candidates, it often selects more than one winner and it is extremely manipulable. Finding a practical rule that consistently gives the highest veto power to minorities is still an open question.

\medskip

\textbf{Tradeoffs.} We find that ($q,k$)-majority criterion is compatible with various standard desirable properties for voting rules. The only exception is the \emph{second-order positional dominance} criterion. For each $q<2k/(3k+1)$, Theorem~\ref{theorem-tradeoff} shows that each rule that satisfies the second-order positional dominance criterion does not satisfy the ($q,k$)-majority criterion.

\medskip

\textbf{Condorcet and Dodgson.} Our results highlight the distinction between different Condorcet extensions. Most of the Condorcet-consistent rules satisfy the mutual majority criterion (see footnote \ref{footnote-MM-rules}), and we study several important exceptions: Condorcet least-reversal, and the methods of Black, Young, Simpson, and Dodgson.

One specific open question arises from the incomplete result regarding \emph{Dodgson's rule}: in contrast to other results, Theorem~\ref{theorem_Dodgson} does not specify the tight bound on the quota. The value of the tight bound seems to be a hard question, as  Dodgson's rule is known to be difficult to work with \citep{Bartholdi89,Caragiannis16,Hemaspaandra97}. It is not easy to check whether the profile in Table~\ref{profile_tight_bounds} (which we use for the proofs in the Appendix) gives the worst case for each candidate in the group of mutually supported candidates and, at the same time, the best case for some other candidate outside the group.

\medskip

\textbf{Limitations of our approach.} Our criteria do not provide a comparison for voting rules that satisfy the mutual majority criterion: all such rules have the same high level of majority power and veto power. In fact, all our results for the instant-runoff voting hold for an arbitrary voting rule that satisfies the mutual majority criterion. Designing a proper quantitative criteria that would distinguish these rules is an open question. 

Another question is generalizing our criteria of majority power and veto power to the framework of \emph{multi-winner elections}, as, for instance, done to a proportionality degree in \cite{LacknerSkowron18} and \cite{Skowron18}. Similarly, an important open question is whether analogous criteria for grading systems--such as approval voting \citep{Brams09} as well as majority judgment \citep{BalinskiLaraki11}--should be developed.

A more general open question is the analysis of majority power and veto power in practically-relevant scenarios. In this paper the main results are based on the worst-case analysis as it allows us to provide precise estimates for any total number of candidates. Future research can make use of more realistic scenarios inspired by theories of individual decision-making, empirical results, and experiments on voting. A particularly developed approach is the one that measures the statistical properties of voting rules, such as Condorcet efficiency \citep{GehrleinLepelley17}, majority winner and majority loser efficiency \citep{Diss18}, manipulability \citep{AleskerovKurbanov99,Aleskerov12,Green-Armytage16}, and others.

In studying majority power and veto power one is not restricted to \emph{single-winner elections} where a representative or a ruler is selected. Alternative ways to protect minorities range from using two periods of voting \citep{FahrenbergerGersbach10}, allowing storable votes in multi-issue elections \citep{Casella05}, direct democracy, participatory budgeting \citep{Cabannes04}, and multi-winner elections. 

\medskip

\textbf{Resume.} As a final takeaway, our results suggest that societies that care about the rights of the majority to make their most preferred candidates win and veto their least preferred candidates should adopt the instant-runoff voting. In contrast, societies that care about the rights of minorities should select Black's rule.

\bigskip

\noindent {\bf Appendix}\nopagebreak

\medskip

\noindent\textbf{Proof of Theorem~\ref{theorem_plurality}}.\nopagebreak

Let $m\geq 3$, and let more than $nk/(k+1)$ voters give candidates
from some subset $B\subsetneq A$ ($m>|B|=k\geq 1$) top $k$
positions. Then, together, all candidates in $B$ receive strictly more
than $n k/(k+1)$ top positions, while candidates from $A\setminus B$
all together receive strictly less than $n/(k+1)$ top positions.
Therefore, at least one of the candidates in $B$ receives strictly
more than $n/(k+1)$ top positions, and each candidate from
$A\setminus B$ receives strictly less than $n/(k+1)$ top positions.
Therefore, the plurality rule can only select a candidate
from set~$B$.

For any smaller quota $q<k/(k+1)$ we can always find the following
counterexample. Let the total number of voters be $n=k+1$ and let
$k$ voters give candidates from set $B$ top $k$ positions such that
each of these candidates gets the top position exactly once. Let the other voter give the top position to some other candidate $a \notin
B$. Then, the plurality rule selects all candidates from the
set $B\cup {a}$.

\medskip

\noindent\textbf{Proof of Theorem~\ref{theorem_Simpson}}.\nopagebreak

Let $m \geq 3$, and more than $n(k-1)/k$ voters top-rank $k\geq 2$
candidates, denote this subset of candidates as $B = \{b_1, \ldots,
b_k\}$. It is easy to see that each candidate in $A\setminus B$ gets
less than $n/k$ of Simpson's scores (a candidate from $A\setminus B$
gets the highest score when it is top-ranked by all voters that do
not top-rank $B$).

Denote the number of first positions of some candidate $b\in B$
among all other candidates in $B$ as $n_1(b,B)$:
\begin{equation}\label{n_1b}
n_1(b,B) = |\{ i: b\succ_i b' \quad \mbox{for all} \quad b'\in
B\setminus b \}|.
\end{equation}

Since the total number of first positions is fixed $n_1(b_1, B) +
\ldots + n_1(b_k, B) = n$, there is a candidate $b\in B$ with a weakly higher number of top positions than average:
$n_1(b,B)\geq n/k$.

Hence, there is a candidate that receives not less than $n/k$ of
scores, and each candidate from $A\setminus B$ gets less than $n/k$
scores and cannot be the winner.

To see that the bound $q=(k-1)/k$ is tight consider the following
counterexample in Table~\ref{profile_tight_bounds}: each candidate $b \in B$ receives exactly
$qn/k$ first positions, $qn/k$ second positions, and so on from the
qualified mutual majority of $qn$ voters, while all voters outside the qualified mutual majority top-rank some other candidate $a_1$ and also prefer all candidates in $A\setminus B$ over candidates in~$B$.

\begin{table}[ht!]
\caption{Preference profile}
\label{profile_tight_bounds}
\begin{tabular}{|c|c|c|c|c|c|}
\hline
$\frac{qn}{k}$ & $\ldots$ & $\frac{qn}{k}$ & $\frac{(1-q)n}{k}$ & $\ldots$ & $\frac{(1-q)n}{k}$  \\
\hline
$b_1$ & $\ldots$ & $b_k$ & $a_1$ & $\ldots$ & $a_1$ \\
$b_2$ & $\ldots$ & $b_1$ & $\ldots$ & $\ldots$ & $\ldots$ \\
$\ldots$ & $\ldots$ & $\ldots$ & $a_{m-k}$ & $\ldots$ & $a_{m-k}$ \\
$b_k$ & $\ldots$ & $b_{k-1}$ & $b_1$ & $\ldots$ & $b_{k}$ \\
$a_1$ & $\ldots$ & $a_1$ & $\ldots$ & $\ldots$ & $\ldots$ \\
$\ldots$ & $\ldots$ & $\ldots$ & $b_{k-1}$ & $\ldots$ & $b_{k-2}$ \\
$a_{m-k}$ & $\ldots$ & $a_{m-k}$ & $b_k$ & $\ldots$ & $b_{k-1}$ \\
\hline
\end{tabular}

\vspace{0.2cm}

\justify
\footnotesize{\emph{Notes}: The qualified mutual majority of $qn$ voters give exactly $qn/k$ first, second, and so on positions to each candidate $b_i \in B$, all preferences over remaining alternatives $A\setminus B$ are the same. The other $(1-q)n$ voters prefer each candidate in $A\setminus B$ over each candidate in $B$, but have a relative orderings within these two sets $B$ and $A \setminus B$ that is identical to the former $qn$ voters. This type of cyclical preferences over $B$ is known as a Condorcet k-tuple.}
\end{table}

For each $k{>}1$ we can set $n{=}k^2$ and $q{=}(k{-}1)/k$. Then, set~$B$ is supported by $n(k{-}1)/k$ voters, while each candidate from the set $B\cup a_1$ gets the same Simpson's score.

\medskip

\noindent\textbf{Proof of Theorem~\ref{theorem_Young}}.\nopagebreak


Let $m \geq 3$, and let more than $n(k-1)/k$ voters top-rank $k\geq 2$ candidates, and denote this subset of candidates as $B = \{b_1, \ldots, b_k\}$. We consider the next two cases separately.

\emph{Case 1}. $[n/k]=n/k$, where $[~]$ is the integer part.

In this case, not more than $[n/k]-1$ voters give the first positions to the candidates from the set $A\setminus B$. For each candidate from $A\setminus B$ to make him a weak Condorcet winner, we need to remove at least $n-2([n/k]-1)=n-2[n/k]+2$ voters.

Consider some candidate $b\in B$ with a higher than average number
of top positions $n_1(b,B)\geq n/k = [n/k]$ (as defined in equation
(\ref{n_1b})). For $b$ to win, at most $n-2[n/k]$ voters have to be removed.

\emph{Case 2}. $[n/k]<n/k$, where $[~]$ is the integer part.

In this case, not more than $[n/k]$ voters give the first positions to the candidates from the set $A\setminus B$. For each candidate from $A\setminus B$ to make him a weak Condorcet winner, we need to remove at least $n-2[n/k]$ voters.

Consider some candidate $b\in B$ with a higher than average number of top positions $n_1(b,B)\geq n/k > [n/k]$ (as defined in equation (\ref{n_1b})). Thus, $n_1(b,B) \geq [n/k]+1$. For $b$ to win, at most $n-2([n/k]+1)=n-2[n/k]-2$ voters have to be removed.

The example from Table~\ref{profile_tight_bounds} shows that the bound $(k-1)/k$ is tight.

\medskip

\noindent\textbf{Proof of Theorem~\ref{theorem_CLR}}.\nopagebreak

Again, we use the preference profile in Table~\ref{profile_tight_bounds}. Let us first show that it is the worst possible profile for each candidate $b \in B$ to win by the Condorcet least-reversal rule, i.e., it has the maximum minimal score $p^{CLR}_b$ among all candidates $b\in B$. To maximize the minimal score $p^{CLR}$ for candidates in $B$ we can maximize the scores (\ref{Clr}) for the subset $B$ separately: $\sum\limits_{c\in B\setminus b}$. This is true, because the other part $\sum\limits_{c\in A\setminus B}$ is zero whenever $q\geq 1/2$.

According to Proposition 5 in \citet{Saari00}, each tournament matrix
with $k$ candidates has unique representation as the sum of its
transitive matrix and its Condorcet k-tuple matrix (Table~\ref{profile_k_tuple}).\footnote{Tournament matrix $h$ is called \emph{transitive} if there exists a linear order $\succ_0\in L(A)$ such that $h(a,b)\geq h(b,a)$ whenever
$a\succ_0 b$. For example, the tournament matrix in Table~\ref{profile_main} is transitive with the linear order $a\succ_0 b\succ_0 c\succ_0 d$.}

\begin{table}[ht!]
\caption{Condorcet k-tuple profile and tournament matrix}
\label{profile_k_tuple}
\begin{tabular}{|c|c|c|c|}
\hline
$\frac{n}{k}$ & $\frac{n}{k}$ & $\ldots$ & $\frac{n}{k}$  \\
\hline
$b_1$ & $b_2$ & $\ldots$ & $b_k$ \\
$b_2$ & $b_3$ & $\ldots$ & $b_1$ \\
$\ldots$ & $\ldots$ & $\ldots$ & $\ldots$ \\
$b_k$ & $b_1$ & $\ldots$ & $b_{k-1}$ \\
\hline
\end{tabular}
\quad
\begin{tabular}{|c|c|c|c|c|c|c|}
\hline
  & $b_1$ & $b_2$ & $\ldots$ & $b_k$   \\
\hline
$b_1$ &   &   $n(k-1)/k$ & $\ldots$ & $n/k$  \\
\hline
$b_2$ &  $n/k$ &   & $\ldots$ & $2n/k$  \\
\hline
$\ldots$ &  $\ldots$ & $\ldots$ & $\ldots$ & $\ldots$  \\
\hline
$b_k$ & $n(k-1)/k$ & $n(k-2)/k$ &  $\ldots$ &    \\
\hline
\end{tabular}
\end{table}

Thus, the maximal element of the transitive matrix gets no more than $p^{CLR}$ total scores in the k-tuple matrix only. Hence, the profile in Table~\ref{profile_tight_bounds} qualifies as the worst case.

Next, we find the bound for the profile in Table~\ref{profile_tight_bounds}. Each candidate
$a\in A\setminus B$ gets at least $p^{CLR}_a\geq nk(2q-1)/2$ points.

Candidate $b_1$ gets $n/k, 2n/k, \ldots, (k-1)n/k$ pairwise majority
wins against candidates ${b_k,\ldots,b_{2}}$ correspondingly. For
even $k$ the score for each $b\in B$ is $p^{CLR}_b=n(k-2)/8$, for
odd $k$ the score is $p^{CLR}_b=n(k-1)^2/(8k)$. Setting these scores
equal to the score $p^{CLR}_{a_{1}}=nk(2q-1)/2$ received by $a_1$ we
get the tight bounds.

\medskip

\noindent\textbf{Proof of Theorem~\ref{theorem_RV}}.\nopagebreak

In case $k=m-1$ and $q=1/2$, in the second round there is at least one candidate from the supported $k$ candidates, and this candidate wins. For any smaller quota $q<1/2$ we can always construct a counterexample where a majority winner does not belong to the set of $k$ candidates. This case also includes the case $m=3, k=2$.

Let $m > 3$, and let more than $nk/(k+2)$ voters give candidates from some subset $B\subsetneq A$ ($m-1>|B|=k > 1$) top $k$ positions. Then, all the candidates in $B$ receive strictly more than $n k/(k+2)$ top positions, while the candidates from $A\setminus B$ all receive strictly less than $2n/(k+2)$ top positions.
Therefore, at least one of the candidates in $B$ and at most one of
the candidates in $A\setminus B$ receive strictly more than
$n/(k+2)$ of top positions. Thus, in the second round there is at
least one candidate from set $B$. Even if the second candidate is
from $A\setminus B$, this second candidate loses to the candidate
from $B$ by simple majority rule. Hence, the winner is from~$B$.

For any smaller quota $q<k/(k+2)$ we can always find the following
counterexample. Let the total number of voters be $n=(k+2)n'+2$ and
let $k n'$ voters give $k$ candidates from set $B$ top $k$ positions
such that each candidate in $B$ gets the top position exactly $n'$
times. Consider the other $2\cdot(n'+1)$ voters and two other
candidates $a_1,a_2 \notin B$. Let $n'+1$ voters top-rank candidate
$a_1$ and the other $n'+1$ voters top-rank candidate $a_2$. Then,
candidates $a_1$ and $a_2$ make it to the second round.

If we set $n' > 2q/(k-kq-2q)$ then set $B$ is supported by more than
$q n$ voters.

\medskip

\noindent\textbf{Proof of Theorem~\ref{theorem_Black}}.\nopagebreak

Let $m\geq 3$, and let more than $qn$ voters give candidates from
some subset $B\subsetneq A$ ($m>|B|=k> 1$) top $k$ positions. Then
one can find the tight bound for the quota $q=q(k,m)$ using the
following equation:

\begin{equation*}
(1-q)(m-1)+q(m-k-1) = q \frac{m-1+m-k}{2} + (1-q) \frac{k-1}{2},
\end{equation*}
where the left part is the maximal Borda score for any $a\notin B$,
and the right part is the minimal possible value for a maximal Borda score within the set~$B$.

The example from Table~\ref{profile_tight_bounds} shows that the bound $(2m-k-1)/(2m)$ is tight.

\medskip

\noindent\textbf{Proof of Theorem~\ref{theorem_convex_median}}.\nopagebreak

Let $qn$ voters give candidates from some subset $B=\{b_1,\ldots,b_k\}$ top $k$ positions. Then, each candidate
$a\notin B$ gets the following truncated Borda score with $t = 2kq$:

\begin{equation*}
\frac{B_{2kq}(a)}{2kq} \leq (1-q)n + \frac{(2kq-k)qn}{2kq}  =
\frac{n}{2}.
\end{equation*}

Let $m>2k$ and $q>(3k-1)/(4k)$. It is sufficient to show that for
some $b\in B$ its truncated Borda score is higher:
$B_{2kq}(b)/(2kq)>n/2$. For a contradiction assume that the truncated Borda score with $t = 2kq$ satisfies the following inequality:
\begin{equation*}
\frac{B_{2kq}(b)}{2kq} \leq \frac{n}{2} \quad \mbox{for each} \quad
b\in B.
\end{equation*}

Then,
\begin{equation*}
\frac{(2kq) n_1(b)+\ldots+(2kq-k+1)n_k(b)}{2kq} \leq \frac{n}{2}
\quad \mbox{for each} \quad b\in B,
\end{equation*}
whence, after summing up $k$ inequalities, we get:
\begin{equation*}
\frac{qnk(4kq-k+1)}{4kq} \leq \frac{nk}{2}.
\end{equation*}

The latter inequality contradicts the assumption $q>(3k-1)/(4k)$.

To show that the bound is tight we use the preference profile
from Table~\ref{profile_tight_bounds}.

Similarly, we find a tight bound for the case $2k\geq m \geq k+1$:
\begin{equation*}
\min_{\succ} \max_{b\in B} \frac{B_{2kq}(b)}{2kq} =
\frac{\frac{qn}{k}\frac{(4kq-k+1)}{2}k}{2kq} +
\frac{\frac{(1-q)n}{k}\frac{(4kq-m)}{2}(2k-m+1)}{2kq} = \frac{n}{2},
\end{equation*}
which leads to equation (\ref{ThCM}) and also to a special case where $m=k+1, q=1/2$.

\medskip

\noindent\textbf{Proof of Theorem~\ref{theorem_Dodgson}}.\nopagebreak

Let more than $k/(k+1)$ of voters give candidates from some subset $B$ top $k$ positions. Then, each candidate $a\notin B$ gets less than $n/(k+1)$ votes in a pairwise comparison to each candidate from the set $B$. Upgrading candidate $a$ by one position in the preference profile adds not more than one vote in a pairwise comparison to each candidate from the set $B$. Therefore, candidate $a$ needs more than $k(\frac{n}{2}-\frac{n}{k+1})$ upgrades to become a Condorcet winner. A candidate from $B$ that gets more than $n/(k+1)$ top positions does not need more than $(k-1)(\frac{n}{2}-\frac{n}{k+1})$ upgrades in the preference
profile in order to become a Condorcet winner. Since
$(k-1)(\frac{n}{2}-\frac{n}{k+1})<k(\frac{n}{2}-\frac{n}{k+1})$,
Dodgson's rule selects from set~$B$.

The second statement follows from the calculations for the profile in Table~\ref{profile_tight_bounds}.

\medskip

\noindent\textbf{Proof of Theorem \ref{theorem_PVC}}.\nopagebreak

From the definition of proportional veto core it follows that for each $k,m$ and $q=(m-k)/m$ this rule satisfies the $(q,k,m)$-majority criterion.  

We prove by contradiction. Assume that the rule satisfies the criterion for some $k,m$ and $q<(m-k)/m$. Then, we can always construct a profile (because the number of voters is arbitrary) such that $0<\varepsilon< -q+(m-k)/m$ and:\\
i) $-\varepsilon + (m-k)/m$ of voters give to some candidates $\{b_1,\ldots,b_k\}$ the highest $k$ positions;\\
ii) other $\varepsilon + k/m$ of voters give to these $k$ candidates the lowest $k$ positions.  

Condition i) and the assumption imply that the choice set is included in the set $\{b_1,\ldots,b_k\}$. Condition ii) implies that none of these $k$ candidates can win. Hence, the assumption is false and the bound $q=(m-k)/m$ is tight.

\medskip

\noindent\textbf{Proof of Theorem~\ref{theorem_scoring}}.\nopagebreak

Let $m\geq 3$, and let more than $qn$ voters give candidates from
some subset $B\subsetneq A$ ($m>|B|=k \geq 1$) top $k$ positions.
One can then find the tight bound for the quota $q=q(k,s_1,\ldots,
s_m)$ using the following equation:

\begin{equation*}
(1-q)\cdot s_1 + q\cdot s_{k+1} = q\cdot  \frac{s_1 + \ldots +
s_k}{k} + (1-q)\cdot \frac{s_m + \ldots + s_{m-k+1}}{k},
\end{equation*}
where the left part is the maximal total score for any $a\notin B$,
and the right part is the minimal maximal total score for any $b\in
B$.

The example from Table~\ref{profile_tight_bounds} shows that the bound (\ref{qMSR}) is tight.

In particular, in case $k=m-1$ and $q=1/2$, we get the inequality (\ref{ML-inequality}).

\medskip

\noindent\textbf{Proof of Theorem~\ref{theorem-tradeoff}}.\nopagebreak

1) Fix any $k \geq 1$ and any $q < 2k/(3k+1)$. To show that the properties 2-PD and ($q,k$)-majority criterion are incompatible we use the profile from Table~\ref{profile_tight_bounds}. 

Let $m > 2k$ and the candidates from the set $B=\{ b_1, \ldots, b_k \}$ constitute a qualified mutual majority of $2k/(3k+1)-\varepsilon$ with a small $\varepsilon>0$. Let us show that candidate $a_1\notin B$ positionally dominates each candidate~$b\in B$ according to the second-order positional dominance. It is equivalent to the inequalities $B_t(a_1) \geq B_t(b)$ for each $t=1,\ldots,m-1$, and $B_{m-1}(a_1) > B_{m-1}(b)$ \citep[see][]{Stein94,Kondratev18}. In our case, it leads to the next obvious inequalities:
\begin{equation*}
    t\left(1-\frac{2k}{3k+1}+\varepsilon\right)> \frac{t(t+1)}{2k}\left(\frac{2k}{3k+1}-\varepsilon\right), \quad 1\leq t\leq k,
\end{equation*}
\begin{multline*}
    t\left(1-\frac{2k}{3k+1}+\varepsilon\right) + (t-k)\left(\frac{2k}{3k+1}-\varepsilon\right)> \frac{2t-k+1}{2}\left(\frac{2k}{3k+1}-\varepsilon\right), \\ \quad k+1\leq t\leq m-k,
\end{multline*}
\begin{multline*}
    t\left(1-\frac{2k}{3k+1}+\varepsilon\right) + (t-k)\left(\frac{2k}{3k+1}-\varepsilon\right)> \frac{2t-k+1}{2}\left(\frac{2k}{3k+1}-\varepsilon\right) \\ + \frac{(t-m+k+1)(t-m+k)}{2k} \left(1-\frac{2k}{3k+1}+\varepsilon\right), \quad m-k+1\leq t\leq m-1.
\end{multline*}

2) Let us construct a rule that satisfies the 2-PD and ($q,k$)-majority criterion for each $k \geq 1$ and each $q \geq 2k/(3k+1)$. 

If $n_1(a)>n/2$ for some candidate~$a$ then this candidate is the winner. Otherwise, for each candidate~$a$ define the next score:
\begin{equation}\label{score-tradeoff}
     \max\left\{ t \in [1, \infty] : \frac{(3t+1)}{2(t+1)} \frac{B_t(a)}{t} \leq \frac{n}{2} \right\},
\end{equation}
and the winner is the candidate with the lowest score. Note that the functions $(3t+1)/(2(t+1))$ and $B_t(a)/t$ are non-decreasing.

Let more than $2k/(3k+1)$ of voters give candidates from some subset
$B=\{b_1,\ldots,b_k\}$ top $k$ positions. Then, we have the next inequality:
\begin{equation*}
\frac{(3k+1)}{2(k+1)} \frac{B_k(a)}{k} < \frac{n}{2}  \quad \mbox{for each} \quad a\notin B,
\end{equation*}
hence, for each $a\notin B$, the score~(\ref{score-tradeoff}) is higher than~$k$.

It is sufficient to show that for some $b\in B$ its score~(\ref{score-tradeoff}) is lower than~$k$. For a contradiction, assume that we have the following inequality:
\begin{equation*}
\frac{(3k+1)}{2(k+1)} \frac{B_k(b)}{k} \leq \frac{n}{2} \quad \mbox{for each} \quad b\in B.
\end{equation*}

Then
\begin{equation*}
\frac{(3k+1)}{2(k+1)} \frac{k n_1(b)+ (k-1) n_2(b) + \ldots+n_k(b)}{k} \leq \frac{n}{2}
\quad \mbox{for each} \quad b\in B,
\end{equation*}
whence, after summing up $k$ inequalities and using the fact that 
\begin{equation*}
    \sum\limits_{b\in B}{n_j(b)>\frac{2kn}{3k+1}} \quad \mbox{for each} \quad j=1,\ldots,k,
\end{equation*}
we receive a contradiction.

\bibliographystyle{apalike}
\bibliography{my_bibliography}

\end{document}